EUROPEAN ORGANIZATION FOR NUCLEAR RESEARCH

# A Search for $\eta_c'$ Production in Photon-Photon Fusion at LEP

DELPHI Collaboration

## Abstract

A search for the production of the $\eta_c'$ meson, the first radial excitation of the ground state of charmonium $\eta_c(2980)$, in the photon-photon fusion reaction at LEP has been performed using the data collected by the DELPHI detector during 1992-1996. No evidence of $\eta_c'$ production is found in the mass region 3520–3800 MeV/c$^2$. An upper limit for the ratio of the two-photon widths of the $\eta_c'$ and $\eta_c$ is obtained.

P.Abreu[21], W.Adam[50], T.Adye[37], P.Adzic[11], G.D.Alekseev[16], R.Alemany[49], P.P.Allport[22], S.Almehed[24], U.Amaldi[9], S.Amato[47], E.G.Anassontzis[3], P.Andersson[44], A.Andreazza[9], S.Andringa[21], P.Antilogus[25], W-D.Apel[17], Y.Arnoud[14], B.Åsman[44], J-E.Augustin[25], A.Augustinus[9], P.Baillon[9], P.Bambade[19], F.Barao[21], G.Barbiellini[46], R.Barbier[25], D.Y.Bardin[16], G.Barker[9], A.Baroncelli[39], M.Battaglia[15], M.Baubillier[23], K-H.Becks[52], M.Begalli[6], P.Beilliere[8], Yu.Belokopytov[9,53], K.Belous[42], A.C.Benvenuti[5], C.Berat[14], M.Berggren[25], D.Bertini[25], D.Bertrand[2], M.Besancon[40], F.Bianchi[45], M.Bigi[45], M.S.Bilenky[16], M-A.Bizouard[19], D.Bloch[10], H.M.Blom[30], M.Bonesini[27], W.Bonivento[27], M.Boonekamp[40], P.S.L.Booth[22], A.W.Borgland[4], G.Borisov[19], C.Bosio[39], O.Botner[48], B.Bouquet[19], C.Bourdarios[19], T.J.V.Bowcock[22], I.Boyko[16], I.Bozovic[11], M.Bozzo[13], P.Branchini[39], T.Brenke[52], R.A.Brenner[48], P.Bruckman[18], J-M.Brunet[8], L.Bugge[33], T.Buran[33], T.Burgsmueller[52], P.Buschmann[52], S.Cabrera[49], M.Caccia[27], M.Calvi[27], A.J.Camacho Rozas[41], T.Camporesi[9], V.Canale[38], F.Carena[9], L.Carroll[22], C.Caso[13], M.V.Castillo Gimenez[49], A.Cattai[9], F.R.Cavallo[5], Ch.Cerruti[10], V.Chabaud[9], Ph.Charpentier[9], L.Chaussard[25], P.Checchia[36], G.A.Chelkov[16], R.Chierici[45], P.Chochula[7], V.Chorowicz[25], J.Chudoba[29], P.Collins[9], M.Colomer[49], R.Contri[13], E.Cortina[49], G.Cosme[19], F.Cossutti[40], J-H.Cowell[22], H.B.Crawley[1], D.Crennell[37], G.Crosetti[13], J.Cuevas Maestro[34], S.Czellar[15], G.Damgaard[28], M.Davenport[9], W.Da Silva[23], A.Deghorain[2], G.Della Ricca[46], P.Delpierre[26], N.Demaria[9], A.De Angelis[9], W.De Boer[17], S.De Brabandere[2], C.De Clercq[2], B.De Lotto[46], A.De Min[36], L.De Paula[47], H.Dijkstra[9], L.Di Ciaccio[38], A.Di Diodato[38], J.Dolbeau[8], K.Doroba[51], M.Dracos[10], J.Drees[52], M.Dris[31], A.Duperrin[25], J-D.Durand[25,9], R.Ehret[17], G.Eigen[4], T.Ekelof[48], G.Ekspong[44], M.Ellert[48], M.Elsing[9], J-P.Engel[10], B.Erzen[43], M.Espirito Santo[21], E.Falk[24], G.Fanourakis[11], D.Fassouliotis[11], J.Fayot[23], M.Feindt[17], A.Fenyuk[42], P.Ferrari[27], A.Ferrer[49], E.Ferrer-Ribas[19], S.Fichet[23], A.Firestone[1], P.-A.Fischer[9], U.Flagmeyer[52], H.Foeth[9], E.Fokitis[31], F.Fontanelli[13], B.Franek[37], A.G.Frodesen[4], R.Fruhwirth[50], F.Fulda-Quenzer[19], J.Fuster[49], A.Galloni[22], D.Gamba[45], S.Gamblin[19], M.Gandelman[47], C.Garcia[49], J.Garcia[41], C.Gaspar[9], M.Gaspar[47], U.Gasparini[36], Ph.Gavillet[9], E.N.Gazis[31], D.Gele[10], J-P.Gerber[10], N.Ghodbane[25], I.Gil[49], F.Glege[52], R.Gokieli[51], B.Golob[43], P.Goncalves[21], I.Gonzalez Caballero[41], G.Gopal[37], L.Gorn[1,54], M.Gorski[51], Yu.Gouz[42], V.Gracco[13], J.Grahl[1], E.Graziani[39], C.Green[22], P.Gris[40], K.Grzelak[51], M.Gunther[48], J.Guy[37], F.Hahn[9], S.Hahn[52], S.Haider[9], A.Hallgren[48], K.Hamacher[52], F.J.Harris[35], V.Hedberg[24], S.Heising[17], J.J.Hernandez[49], P.Herquet[2], H.Herr[9], T.L.Hessing[35], J.-M.Heuser[52], E.Higon[49], S-O.Holmgren[44], P.J.Holt[35], D.Holthuizen[30], S.Hoorelbeke[2], M.Houlden[22], J.Hrubec[50], K.Huet[2], K.Hultqvist[44], J.N.Jackson[22], R.Jacobsson[9], P.Jalocha[9], R.Janik[7], Ch.Jarlskog[24], G.Jarlskog[24], P.Jarry[40], B.Jean-Marie[19], E.K.Johansson[44], P.Jonsson[24], C.Joram[9], P.Juillot[10], F.Kapusta[23], K.Karafasoulis[11], S.Katsanevas[25], E.C.Katsoufis[31], R.Keranen[17], Yu.Khokhlov[42], B.A.Khomenko[16], N.N.Khovanski[16], A.Kiiskinen[15], B.King[22], N.J.Kjaer[30], O.Klapp[52], H.Klein[9], P.Kluit[30], D.Knoblauch[17], P.Kokkinias[11], A.Konopliannikov[42], M.Koratzinos[9], V.Kostioukhine[42], C.Kourkoumelis[3], O.Kouznetsov[16], M.Krammer[50], C.Kreuter[9], E.Kriznic[43], J.Krstic[11], Z.Krumstein[16], P.Kubinec[7], W.Kucewicz[18], K.Kurvinen[15], J.W.Lamsa[1], D.W.Lane[1], P.Langefeld[52], V.Lapin[42], J-P.Laugier[40], R.Lauhakangas[15], G.Leder[50], F.Ledroit[14], V.Lefebure[2], L.Leinonen[44], A.Leisos[11], R.Leitner[29], J.Lemonne[2], G.Lenzen[52], V.Lepeltier[19], T.Lesiak[18], M.Lethuillier[40], J.Libby[35], D.Liko[9], A.Lipniacka[44], I.Lippi[36], B.Loerstad[24], M.Lokajicek[12], J.G.Loken[35], J.H.Lopes[47], J.M.Lopez[41], R.Lopez-Fernandez[14], D.Loukas[11], P.Lutz[40], L.Lyons[35], J.R.Mahon[6], A.Maio[21], A.Malek[52], T.G.M.Malmgren[44], V.Malychev[16], F.Mandl[50], J.Marco[41], R.Marco[41], B.Marechal[47], M.Margoni[36], J-C.Marin[9], C.Mariotti[9], A.Markou[11], C.Martinez-Rivero[19], F.Martinez-Vidal[49], S.Marti i Garcia[22], N.Mastroyiannopoulos[11], F.Matorras[41], C.Matteuzzi[27], G.Matthiae[38], J.Mazik[29], F.Mazzucato[36], M.Mazzucato[36], M.Mc Cubbin[22], R.Mc Kay[1], R.Mc Nulty[9], G.Mc Pherson[22], C.Meroni[27], W.T.Meyer[1], A.Miagkov[42], E.Migliore[45], L.Mirabito[25], W.A.Mitaroff[50], U.Mjoernmark[24], T.Moa[44], R.Moeller[28], K.Moenig[9], M.R.Monge[13], X.Moreau[23], P.Morettini[13], G.Morton[35], U.Mueller[52], K.Muenich[52], M.Mulders[30], C.Mulet-Marquis[14], R.Muresan[24], W.J.Murray[37], B.Muryn[14,18], G.Myatt[35], T.Myklebust[33], F.Naraghi[14], F.L.Navarria[5], S.Navas[49], K.Nawrocki[51], P.Negri[27], N.Neufeld[9], N.Neumeister[50], R.Nicolaidou[14], B.S.Nielsen[28], V.Nikolaenko[10], M.Nikolenko[10,16], V.Nomokonov[15], A.Normand[22], A.Nygren[24], A.Oblakowska-Mucha[18], V.Obraztsov[42], A.G.Olshevski[16], A.Onofre[21], R.Orava[15], G.Orazi[10], K.Osterberg[15], A.Ouraou[40], M.Paganoni[27], S.Paiano[5], R.Pain[23], R.Paiva[21], J.Palacios[35], H.Palka[18], Th.D.Papadopoulou[31], K.Papageorgiou[11], L.Pape[9], C.Parkes[35], F.Parodi[13], U.Parzefall[22], A.Passeri[39], O.Passon[52], M.Pegoraro[36], L.Peralta[21], M.Pernicka[50], A.Perrotta[5], C.Petridou[46], A.Petrolini[13], H.T.Phillips[37], G.Piana[13], F.Pierre[40], E.Piotto[27], T.Podobnik[43], M.E.Pol[6], G.Polok[18], P.Poropat[46], V.Pozdniakov[16], P.Privitera[38], N.Pukhaeva[16], A.Pullia[27], D.Radojicic[35], S.Ragazzi[27], H.Rahmani[31], D.Rakoczy[50], P.N.Ratoff[20], A.L.Read[33], P.Rebecchi[9], N.G.Redaelli[27], M.Regler[50], D.Reid[9], R.Reinhardt[52], P.B.Renton[35], L.K.Resvanis[3], F.Richard[19], J.Ridky[12], G.Rinaudo[45], O.Rohne[33], A.Romero[45], P.Ronchese[36], E.I.Rosenberg[1], P.Rosinsky[7], P.Roudeau[19], T.Rovelli[5], V.Ruhlmann-Kleider[40], A.Ruiz[41], H.Saarikko[15], Y.Sacquin[40], A.Sadovsky[16], G.Sajot[14], J.Salt[49], D.Sampsonidis[11], M.Sannino[13], H.Schneider[17], Ph.Schwemling[23], U.Schwickerath[17], M.A.E.Schyns[52], F.Scuri[46], P.Seager[20], Y.Sedykh[16], A.M.Segar[35], R.Sekulin[37], Kamal K.Seth[32], R.C.Shellard[6], A.Sheridan[22], M.Siebel[52], R.Silvestre[40], L.Simard[40], F.Simonetto[36], A.N.Sisakian[16], T.B.Skaali[33], G.Smadja[25], O.Smirnova[24], G.R.Smith[37], A.Sopczak[17], R.Sosnowski[51], T.Spassov[21], E.Spiriti[39], P.Sponholz[52], S.Squarcia[13], D.Stampfer[50], C.Stanescu[39], S.Stanic[43], S.Stapnes[33], K.Stevenson[35], A.Stocchi[19], J.Strauss[50], R.Strub[10], B.Stugu[4], M.Szczekowski[51], M.Szeptycka[51], T.Tabarelli[27], F.Tegenfeldt[48], F.Terranova[27], J.Thomas[35], A.Tilquin[26], J.Timmermans[30], L.G.Tkatchev[16], T.Todorov[10], S.Todorova[10], D.Z.Toet[30], A.Tomaradze[32], B.Tome[21], A.Tonazzo[27], L.Tortora[39],

G.Transtromer[24], D.Treille[9], G.Tristram[8], C.Troncon[27], A.Tsirou[9], M-L.Turluer[40], I.A.Tyapkin[16], S.Tzamarias[11], B.Ueberschaer[52], O.Ullaland[9], V.Uvarov[42], G.Valenti[5], E.Vallazza[46], G.W.Van Apeldoorn[30], P.Van Dam[30], J.Van Eldik[30], A.Van Lysebetten[2], I.Van Vulpen[30], N.Vassilopoulos[35], G.Vegni[27], L.Ventura[36], W.Venus[37], F.Verbeure[2], M.Verlato[36], L.S.Vertogradov[16], V.Verzi[38], D.Vilanova[40], L.Vitale[46], E.Vlasov[42], A.S.Vodopyanov[16], G.Voulgaris[3], V.Vrba[12], H.Wahlen[52], C.Walck[44], C.Weiser[17], D.Wicke[52], J.H.Wickens[2], G.R.Wilkinson[9], M.Winter[10], M.Witek[18], G.Wolf[9], J.Yi[1], O.Yushchenko[42], A.Zaitsev[42], A.Zalewska[18], P.Zalewski[51], D.Zavrtanik[43], E.Zevgolatakos[11], N.I.Zimin[16,24], G.C.Zucchelli[44], G.Zumerle[36]

[1]Department of Physics and Astronomy, Iowa State University, Ames IA 50011-3160, USA
[2]Physics Department, Univ. Instelling Antwerpen, Universiteitsplein 1, BE-2610 Wilrijk, Belgium
and IIHE, ULB-VUB, Pleinlaan 2, BE-1050 Brussels, Belgium
and Faculté des Sciences, Univ. de l'Etat Mons, Av. Maistriau 19, BE-7000 Mons, Belgium
[3]Physics Laboratory, University of Athens, Solonos Str. 104, GR-10680 Athens, Greece
[4]Department of Physics, University of Bergen, Allégaten 55, NO-5007 Bergen, Norway
[5]Dipartimento di Fisica, Università di Bologna and INFN, Via Irnerio 46, IT-40126 Bologna, Italy
[6]Centro Brasileiro de Pesquisas Físicas, rua Xavier Sigaud 150, BR-22290 Rio de Janeiro, Brazil
and Depto. de Física, Pont. Univ. Católica, C.P. 38071 BR-22453 Rio de Janeiro, Brazil
and Inst. de Física, Univ. Estadual do Rio de Janeiro, rua São Francisco Xavier 524, Rio de Janeiro, Brazil
[7]Comenius University, Faculty of Mathematics and Physics, Mlynska Dolina, SK-84215 Bratislava, Slovakia
[8]Collège de France, Lab. de Physique Corpusculaire, IN2P3-CNRS, FR-75231 Paris Cedex 05, France
[9]CERN, CH-1211 Geneva 23, Switzerland
[10]Institut de Recherches Subatomiques, IN2P3 - CNRS/ULP - BP20, FR-67037 Strasbourg Cedex, France
[11]Institute of Nuclear Physics, N.C.S.R. Demokritos, P.O. Box 60228, GR-15310 Athens, Greece
[12]FZU, Inst. of Phys. of the C.A.S. High Energy Physics Division, Na Slovance 2, CZ-180 40, Praha 8, Czech Republic
[13]Dipartimento di Fisica, Università di Genova and INFN, Via Dodecaneso 33, IT-16146 Genova, Italy
[14]Institut des Sciences Nucléaires, IN2P3-CNRS, Université de Grenoble 1, FR-38026 Grenoble Cedex, France
[15]Helsinki Institute of Physics, HIP, P.O. Box 9, FI-00014 Helsinki, Finland
[16]Joint Institute for Nuclear Research, Dubna, Head Post Office, P.O. Box 79, RU-101 000 Moscow, Russian Federation
[17]Institut für Experimentelle Kernphysik, Universität Karlsruhe, Postfach 6980, DE-76128 Karlsruhe, Germany
[18]Institute of Nuclear Physics and University of Mining and Metalurgy, Ul. Kawiory 26a, PL-30055 Krakow, Poland
[19]Université de Paris-Sud, Lab. de l'Accélérateur Linéaire, IN2P3-CNRS, Bât. 200, FR-91405 Orsay Cedex, France
[20]School of Physics and Chemistry, University of Lancaster, Lancaster LA1 4YB, UK
[21]LIP, IST, FCUL - Av. Elias Garcia, 14-1$^o$, PT-1000 Lisboa Codex, Portugal
[22]Department of Physics, University of Liverpool, P.O. Box 147, Liverpool L69 3BX, UK
[23]LPNHE, IN2P3-CNRS, Univ. Paris VI et VII, Tour 33 (RdC), 4 place Jussieu, FR-75252 Paris Cedex 05, France
[24]Department of Physics, University of Lund, Sölvegatan 14, SE-223 63 Lund, Sweden
[25]Université Claude Bernard de Lyon, IPNL, IN2P3-CNRS, FR-69622 Villeurbanne Cedex, France
[26]Univ. d'Aix - Marseille II - CPP, IN2P3-CNRS, FR-13288 Marseille Cedex 09, France
[27]Dipartimento di Fisica, Università di Milano and INFN, Via Celoria 16, IT-20133 Milan, Italy
[28]Niels Bohr Institute, Blegdamsvej 17, DK-2100 Copenhagen Ø, Denmark
[29]NC, Nuclear Centre of MFF, Charles University, Areal MFF, V Holesovickach 2, CZ-180 00, Praha 8, Czech Republic
[30]NIKHEF, Postbus 41882, NL-1009 DB Amsterdam, The Netherlands
[31]National Technical University, Physics Department, Zografou Campus, GR-15773 Athens, Greece
[32]Physics Department, Northwestern University, Evanston, IL 60208, USA
[33]Physics Department, University of Oslo, Blindern, NO-1000 Oslo 3, Norway
[34]Dpto. Fisica, Univ. Oviedo, Avda. Calvo Sotelo s/n, ES-33007 Oviedo, Spain
[35]Department of Physics, University of Oxford, Keble Road, Oxford OX1 3RH, UK
[36]Dipartimento di Fisica, Università di Padova and INFN, Via Marzolo 8, IT-35131 Padua, Italy
[37]Rutherford Appleton Laboratory, Chilton, Didcot OX11 OQX, UK
[38]Dipartimento di Fisica, Università di Roma II and INFN, Tor Vergata, IT-00173 Rome, Italy
[39]Dipartimento di Fisica, Università di Roma III and INFN, Via della Vasca Navale 84, IT-00146 Rome, Italy
[40]DAPNIA/Service de Physique des Particules, CEA-Saclay, FR-91191 Gif-sur-Yvette Cedex, France
[41]Instituto de Fisica de Cantabria (CSIC-UC), Avda. los Castros s/n, ES-39006 Santander, Spain
[42]Inst. for High Energy Physics, Serpukov P.O. Box 35, Protvino, (Moscow Region), Russian Federation
[43]J. Stefan Institute, Jamova 39, SI-1000 Ljubljana, Slovenia and Department of Astroparticle Physics, School of Environmental Sciences, Kostanjeviska 16a, Nova Gorica, SI-5000 Slovenia,
and Department of Physics, University of Ljubljana, SI-1000 Ljubljana, Slovenia
[44]Fysikum, Stockholm University, Box 6730, SE-113 85 Stockholm, Sweden
[45]Dipartimento di Fisica Sperimentale, Università di Torino and INFN, Via P. Giuria 1, IT-10125 Turin, Italy
[46]Dipartimento di Fisica, Università di Trieste and INFN, Via A. Valerio 2, IT-34127 Trieste, Italy
and Istituto di Fisica, Università di Udine, IT-33100 Udine, Italy
[47]Univ. Federal do Rio de Janeiro, C.P. 68528 Cidade Univ., Ilha do Fundão BR-21945-970 Rio de Janeiro, Brazil
[48]Department of Radiation Sciences, University of Uppsala, P.O. Box 535, SE-751 21 Uppsala, Sweden
[49]IFIC, Valencia-CSIC, and D.F.A.M.N., U. de Valencia, Avda. Dr. Moliner 50, ES-46100 Burjassot (Valencia), Spain
[50]Institut für Hochenergiephysik, Österr. Akad. d. Wissensch., Nikolsdorfergasse 18, AT-1050 Vienna, Austria
[51]Inst. Nuclear Studies and University of Warsaw, Ul. Hoza 69, PL-00681 Warsaw, Poland
[52]Fachbereich Physik, University of Wuppertal, Postfach 100 127, DE-42097 Wuppertal, Germany
[53]On leave of absence from IHEP Serpukhov
[54]Now at University of Florida

# 1 Introduction

The properties of the charmonium states are well suited for fundamental tests of QCD dynamics [1,2], but they have not yet been well determined. In particular, our understanding of spin-singlet states is extremely poor, being limited to the only well established singlet state in all onium spectroscopy, the $\eta_c(^1S_0)$ ground state of charmonium. The identification of the first radial excitation of the $\eta_c$ was claimed by the Crystal Ball Collaboration in the inclusive photon spectrum of $\psi'$ decays, with a mass and width [3,4] of $m_{\eta_c'} = 3594 \pm 5$ MeV/c$^2$ and $\Gamma(\eta_c') < 8$ MeV. Unfortunately, the state has not been observed in any other experiment. The search for $\eta_c'$ thus poses a worthy challenge to the new ways of studying charmonium spectroscopy, namely $p\bar{p}$ annihilation [2] and photon-photon fusion [5].

Calculations with a variety of $q\bar{q}$ potentials [6], as well as "model independent" calculations based on measured $e^+e^-$ decay widths of $J/\psi$ and $\psi'$ and the well known expression for hyperfine splitting, lead to the prediction that $m_{\eta_c'} = 3615 \pm 10$ MeV/c$^2$. Lattice gauge calculations cannot yet predict the masses of radially excited states accurately, but the first results are consistent with the above prediction [7]. To first order the two photon widths of $\eta_c$ and $\eta_c'$ are proportional to the squares of their respective radial wave functions at the origin [8]. A more elaborate relativistic calculation, which was able to predict $\Gamma_{\gamma\gamma}(\eta_c)$ successfully, predicts that $\Gamma_{\gamma\gamma}(\eta_c')/\Gamma_{\gamma\gamma}(\eta_c) = 0.75 \pm 0.02$ [9][1].

Precision measurements of the properties of charmonium resonances have been made by Fermilab experiments E760 and E835 via their formation in $p\bar{p}$ annihilation [2]. E760/E835 have successfully identified $\eta_c$ in the reaction $p\bar{p} \to \eta_c \to \gamma\gamma$ [10,11], but have failed to find any evidence for $\eta_c'$ in the two photon decay channel. E835 has established a 90% upper confidence limit [11]

$$\frac{Br(p\bar{p} \to \eta_c')Br(\eta_c' \to \gamma\gamma)}{Br(p\bar{p} \to \eta_c)Br(\eta_c \to \gamma\gamma)} \leq 0.16 \tag{1}$$

for an $\eta_c'$ anywhere in the mass region 3570 to 3670 MeV/c$^2$ with $\Gamma(\eta_c') \geq 5$ MeV. Thus the $\eta_c'$ continues to be elusive.

Photon-photon fusion represents another potentially powerful technique for studying positive charge conjugation charmonium resonances $R$, and several attempts to study these in the reaction

$$e^+e^- \to e^+e^-(\gamma\gamma) \to e^+e^-R \to e^+e^-(hadrons) \tag{2}$$

have been reported [12]. Measurements at high collider energies at LEP are especially well suited for these studies because the two photon flux increases with center-of-mass energy $\sqrt{s}$, and the background decreases. Further, given sufficient statistics, several resonances with common decay channels can be investigated in the same invariant mass spectrum. Thus the $\eta_c$ and $\eta_c'$ can be searched for simultaneously.

The present search for $\eta_c'$ in the DELPHI data was motivated by the experimental results mentioned above and by the potential advantages of photon-photon fusion measurements at LEP. Signals for resonance production in the invariant mass range 3520 to 3800 MeV/c$^2$ were searched for in the five decay channels in which $\eta_c$ is known to have large branching ratios for producing charged particles or $K_s^0$: $\rho^0\rho^0$, $K_s^0K^+\pi^-(K_s^0K^-\pi^+)$, $K^{*0}K^-\pi^+(\bar{K}^{*0}K^+\pi^-)$, $K_s^0K_s^0\pi^+\pi^-$ and $K^+K^-K^+K^-$, with $K_s^0 \to \pi^+\pi^-$ and $K^{*0} \to K^+\pi^-$ [4].

[1] The predicted value of the ratio $\Gamma_{\gamma\gamma}(\eta_c')/\Gamma_{\gamma\gamma}(\eta_c) = 0.75 \pm 0.02$ was calculated for $m_{\eta_c'} = 3590$ MeV/c$^2$ [9]. The authors estimate that the effect of varying $m_{\eta_c'}$ over the range $3590 \pm 50$ MeV/c$^2$ is to change the ratio $\Gamma_{\gamma\gamma}(\eta_c')/\Gamma_{\gamma\gamma}(\eta_c)$ by ±4%.

The data used for the analysis were collected with the DELPHI detector at LEP in 1992–1996 with integrated luminosities of 130 $pb^{-1}$ at the $Z^0$ peak, 6 $pb^{-1}$ at 130 and 136 GeV, and 20 $pb^{-1}$ at 161 and 172 GeV.

# 2 Particle selection

A detailed description of the DELPHI detector can be found in [13]. Most of the DELPHI subdetectors were used in the present analysis.

Charged particles were selected if they fulfilled the following criteria:

- polar angle between 20° and 160°;
- momentum greater than 0.4 GeV/c;
- good quality, assessed as follows:
  - track length greater than 50 cm;
  - impact parameters with respect to the nominal interaction point less than 4 cm (transverse and longitudinal with respect to the beam direction);
  - error in momentum measurement less than 100%.

The $K^{\pm}$ identification was based on the ionization measured in the Time Projection Chamber and on the measurement of the angle of emission of its Cherenkov light in the Ring Imaging Cherenkov detectors [14]. The $K^{\pm}$ were selected by imposing the selection criteria in the standard DELPHI algorithm HADSIGN [15].

The $K_s^0$ mesons were detected by their decay in flight into $\pi^+\pi^-$. Such decays are generally separated from the primary vertex, measured for each event. Candidates for secondary decays were found by considering all pairs of particles with opposite charge and applying the selection criteria described in [15].

Other neutral particles were detected from their shower profiles in the High Density Projection Chamber, the Forward Electromagnetic Calorimeters, the Hadron Calorimeter, and the luminosity monitors, SAT and STIC.

# 3 Event selection

The final states $\rho^0\rho^0$, $K_s^0K^+\pi^-$, $K^{*0}K^-\pi^+$, $K_s^0K_s^0\pi^+\pi^-$ and $K^+K^-K^+K^-$, used for this $\eta_c'$ search, contain only charged particles and $K_s^0$ mesons.

The calorimetric information was included in the analysis in order to detect any additional neutral component of the hadronic system and thus provide better rejection of the background. Events containing neutral particles other than $K_s^0$ mesons, which were not associated with charged particles, were removed from the analysis. For the neutral particle identification the same requirements were used as in [16]:

- energy of the electromagnetic or hadron shower greater than 0.5 GeV;
- additional requirements on shower quality, assessed as follows:
  - showers in the luminosity monitor with deposits in more than one cell;
  - error in the hadron calorimeter shower measurement less than 100%.

Only events with net zero charge were used in the following analysis. The vector sum of the transverse momenta of all particles was required to be less than 1.4 GeV/c. The sum of the energy (electromagnetic and hadronic) of all particles was required to be less than 10 GeV.

The selection criteria described above were applied to all channels studied. Additional selection criteria for each channel were applied as follows.

### 3.1 Selection of $\gamma\gamma \to \rho^0\rho^0$ events

Candidate events for the production of $\eta_c'$ decaying into $\rho^0\rho^0 \to \pi^+\pi^-\pi^+\pi^-$ were required to contain just four charged particles, none of which was identified as a charged kaon.

The invariant mass distribution of $\pi^+\pi^-$ pairs for such events is shown in Fig. 1a, where the signal from $\rho^0(770)$ decays is clearly seen. Since the $\rho^0\rho^0$ dynamics largely dominates the $\pi^+\pi^-\pi^+\pi^-$ decay of $\eta_c$ [17], events were selected only if the invariant mass of both $\pi^+\pi^-$ pairs in $\gamma\gamma \to (\pi^+\pi^-)(\pi^+\pi^-)$ events were in the tightly restricted $\rho^0$ mass region of $0.70 < M(\pi\pi) < 0.84$ GeV/c$^2$ for at least one pairing. These cuts are shown in Fig. 1a. The invariant mass distribution of the $\rho^0\rho^0$ pairs in the selected $\gamma\gamma \to \rho^0\rho^0$ events is shown in Fig. 2a.

### 3.2 Selection of $\gamma\gamma \to K_s^0K^\pm\pi^\mp$ events

The selection of $\gamma\gamma \to K_s^0K^\pm\pi^\mp$ events required exactly two charged particles, only one of which was a charged kaon, and one reconstructed $K_s^0$ meson. The invariant mass distribution of the $K_s^0K^\pm\pi^\mp$ in the selected events is shown in Fig. 2b.

### 3.3 Selection of $\gamma\gamma \to K^{*0}K^-\pi^+$ and $\gamma\gamma \to \bar{K}^{*0}K^+\pi^-$ events

Events containing two oppositely charged kaons and two oppositely charged pions were selected to study the channels $\gamma\gamma \to K^{*0}K^-\pi^+$ and $\gamma\gamma \to \bar{K}^{*0}K^+\pi^-$, where the $K^{*0}(890)$ and $\bar{K}^{*0}(890)$ mesons decay into $K^+\pi^-$ and $K^-\pi^+$, respectively. The invariant mass distribution of the two possible $K\pi$ pairs ($K^+\pi^-$ and $K^-\pi^+$) in the events is shown in Fig. 1b. Events were used only if at least one $K\pi$ pair had an invariant mass in the $K^{*0}$ mass region of $0.86 < M(K\pi) < 0.94$ GeV/c$^2$, as indicated in Fig. 1b. The invariant mass distribution of selected final state $K^{*0}K^-\pi^+$ and $\bar{K}^{*0}K^+\pi^-$ particles is shown in Fig. 2c.

### 3.4 Selection of $\gamma\gamma \to K_s^0K_s^0\pi^+\pi^-$ events

The decay channel $K_s^0K_s^0\pi^+\pi^-$ was studied using the events with two $K_s^0$ mesons and two oppositely charged particles, neither of which was identified as a charged kaon. The invariant mass distribution of final state particles in selected $\gamma\gamma \to K_s^0K_s^0\pi^+\pi^-$ events is presented in Fig. 2d.

### 3.5 Selection of $\gamma\gamma \to K^+K^-K^+K^-$ events

The decay channel with four charged kaons $K^+K^-K^+K^-$ in the final state was studied using events with four charged particles of which at least three were identified as charged kaons. The invariant mass distribution of selected final state $K^+K^-K^+K^-$ particles is shown in Fig. 2e.

## 4 Results

As seen in Fig. 2, the invariant mass distributions of final state particles have similar behaviour for all channels; there is an excess of events in the $\eta_c(2980)$ mass region, while no evidence of the $\eta_c'(3594)$ is seen. This is made especially clear in Fig. 3, where the sum of the invariant mass distributions of all five decay channels is shown. There is a clear $\eta_c$ signal above the background and no visible peak near the $\eta_c'(3594)$. In order to quantify these conclusions, the following analysis was performed.

The observed counts $N(m)$ at invariant mass $m$ can be written as the sum of the background counts $N_B(m)$ and the contributions $N_{\eta_c}(m)$ and $N_{\eta_c'}(m)$ of the $\eta_c$ and $\eta_c'$ resonances

$$N(m) = N_B(m) + N_{\eta_c}(m) + N_{\eta_c'}(m)\,. \tag{3}$$

The background counts in Fig. 3 appear to vary linearly with mass, as is indeed expected approximately from the well known two photon flux function $L(s, m)$ [4], where $s$ is the square of the $e^+e^-$ center-of-mass energy. Therefore, $N_B(m)$ was replaced by $a + bm$. The resonance contribution can be written as

$$N_R(m) = \mathcal{L}\ L(s, m)\ \sigma_{BW}(m, m_R, \Gamma_R, \Gamma_{\gamma\gamma}(R))\ Br(R{\to}F)\ \epsilon_F(m)\,. \tag{4}$$

Here $\mathcal{L}$ is the $e^+e^-$ luminosity; $L(s, m)$ is the two photon flux function; $\sigma_{BW}$, which is proportional to $(2J_R + 1)\Gamma_{\gamma\gamma}(R)$, is the Breit-Wigner cross section for the formation of resonance $R$ with mass $m_R$, total width $\Gamma_R$, and two photon width $\Gamma_{\gamma\gamma}(R)$; $Br(R{\to}F)$ is the branching fraction for the decay of the resonance into the final state $F$; and $\epsilon_F(m)$ is the efficiency for detecting this final state.

By passing Monte-Carlo generated events through the full DELPHI detector simulation [15], it was determined that the experimental mass resolution functions for all the decay channels studied here are approximately Gaussian with nearly the same width, $\sigma$=26 MeV and FWHM$\simeq$60 MeV. This width is much larger than the natural width of the $\eta_c$ or $\eta_c'$. This allows eq. (4) to be rewritten as:

$$N_R(m) = \ \alpha_R\ G_R(m, m_R)\,, \tag{5}$$

where $G_R(m, m_R)$ is the Gaussian shape function for resonance $R$ and

$$\alpha_R = \ k\ \mathcal{L}\ L(s, m_R)\ \epsilon_F(m_R)\ Br(R{\to}F)\ (2J_R + 1)\ \Gamma_{\gamma\gamma}(R)\,, \tag{6}$$

where $k$ is a constant.

Thus eq. (3) reduces to

$$N(m) = a + bm + \alpha_{\eta_c}[G_{\eta_c}(m, m_{\eta_c}) + \frac{\alpha_{\eta_c'}}{\alpha_{\eta_c}} G_{\eta_c'}(m, m_{\eta_c'})]\,, \tag{7}$$

where

$$\alpha \equiv \frac{\alpha_{\eta_c'}}{\alpha_{\eta_c}} = \left[\frac{L(\eta_c')}{L(\eta_c)}\right]\left[\frac{\epsilon(\eta_c')}{\epsilon(\eta_c)}\right]\left[\frac{Br(\eta_c'{\to}hadrons)}{Br(\eta_c{\to}hadrons)}\right]\left[\frac{\Gamma_{\gamma\gamma}(\eta_c')}{\Gamma_{\gamma\gamma}(\eta_c)}\right]\,. \tag{8}$$

The invariant mass spectrum $N(m)$ can therefore be fitted with six free parameters: $a$, $b$, $\alpha_{\eta_c}$, $m_{\eta_c}$, $\alpha$ and $m_{\eta_c'}$.

In practice, the value of $m_{\eta_c'}$ was fixed at a series of values covering the range 3520 to 3800 MeV/c$^2$ in 20 MeV/c$^2$ steps, and the best fit values of the remaining five parameters were obtained in each case. The ratio $L(\eta_c')/L(\eta_c)$ was calculated analytically [4] and found to vary from 0.57 to 0.44. The ratio $\epsilon(\eta_c')/\epsilon(\eta_c)$ for the sum of all decay channels studied was determined from Monte-Carlo simulations and found to vary from 1.33 to 1.50. It has been predicted [18,9] that the ratio of the corresponding hadronic decay widths of $\eta_c$ and $\eta_c'$, $\Gamma(\eta_c' \to h_i)/\Gamma(\eta_c \to h_i)$, is a constant[2] and that the total widths of both $\eta_c$ and $\eta_c'$ are nearly 100% hadronic, leading to $Br(\eta_c' \to hadrons)/Br(\eta_c \to hadrons) \simeq 1$. Assuming this to be true for the decay channels used here[3] gives $m_{\eta_c} = 2999 \pm 8$ MeV/c$^2$, a total number of counts in the $\eta_c$ peak of $25.1 \pm 6.4$, and

$$\frac{\Gamma_{\gamma\gamma}(\eta_c')}{\Gamma_{\gamma\gamma}(\eta_c)} = (0.00 \text{ to } 0.06) \pm (0.14 \text{ to } 0.17) \tag{9}$$

for values of $m_{\eta_c'}$ running from 3520 to 3800 MeV/c$^2$. The fit for $m_{\eta_c'}$=3600 MeV/c$^2$ is shown in Fig. 3 by the solid curve.

[2] An analagous prediction about the constancy of the ratio $\Gamma(\psi' \to h_i)/\Gamma(\psi \to h_i)$ has been found to hold for many channels. But there are some striking exceptions, for which there is no generally accepted explanation. The most striking one being the $\rho\pi$ final state, this long-standing problem is often called the "$\rho\pi$ puzzle". For a new explanation of the $\rho\pi$ puzzle, as well as references to earlier work, see [19].

If the data for the different decay channels are analyzed separately, the combined result for the ratio of two photon widths of the $\eta_c'$ and $\eta_c$ is found to be consistent with eq. (9). The error in eq. (9) is only statistical. Systematic errors in the ratio due to variation of the event selection criteria, estimation of efficiency, and the simplifying assumptions made in the analysis are estimated to be $\leq$0.05. Expressed in terms of an upper limit the above result corresponds to

$$\frac{\Gamma_{\gamma\gamma}(\eta_c')}{\Gamma_{\gamma\gamma}(\eta_c)} \leq 0.34\,, \qquad (90\%\ CL). \tag{10}$$

# 5 Discussion

The result of the E835 experiment in eq. (1) can be written as

$$\frac{\Gamma_{\gamma\gamma}(\eta_c')}{\Gamma_{\gamma\gamma}(\eta_c)}\left[\frac{\Gamma_{p\bar{p}}(\eta_c')}{\Gamma_{p\bar{p}}(\eta_c)}\frac{\Gamma^2(\eta_c)}{\Gamma^2(\eta_c')}\right] \leq 0.16\,, \qquad (90\%\ CL). \tag{11}$$

The smallness of this ratio can be due either to the ratio of the hadronic widths in the square bracket or to the ratio of the $\gamma\gamma$ widths. The results in eqs. (9) and (10) indicate that it is the two photon width of the $\eta_c'$ which is much smaller than the two photon width of the $\eta_c$. As mentioned in the introduction, the prediction of a relativistic calculation is that $\Gamma_{\gamma\gamma}(\eta_c')/\Gamma_{\gamma\gamma}(\eta_c) = 0.75 \pm 0.02$. This would correspond to the $\eta_c'$ signal illustrated by the dashed curve in Fig. 3. The 90% confidence limit in eq. (10) is more than a factor two smaller, and it poses an interesting and challenging problem for theory.

# 6 Summary

The first attempt to search for $\eta_c'$ production in two photon reactions has been made using the data collected by the DELPHI detector at LEP during 1992-1996. No evidence of $\eta_c'$ production in $\gamma\gamma$ reactions within the mass region 3520–3800 MeV/c$^2$ has been found. Assuming that, as predicted [18,9], the $\eta_c$ and $\eta_c'$ have the same decay branching ratios, at least for the five channels studied here, the upper limit for the ratio of two-photon widths of $\eta_c'$ and $\eta_c$ is found to be

$$\frac{\Gamma_{\gamma\gamma}(\eta_c')}{\Gamma_{\gamma\gamma}(\eta_c)} \leq 0.34\,, \qquad (90\%\ CL). \tag{12}$$

This limit is essentially determined by event statistics. The high energy running of LEP now in progress is expected to yield a total integrated $e^+e^-$ luminosity of about 500 pb$^{-1}$, with an additional factor of about 1.5 increase in the two photon flux. Thus a factor five increase in the number of events may be hoped for. This should lead to more definitive results for the identification of the $\eta_c'$ and the determination of its two photon width.

[3]Failure of this assumption would presumably indicate that an interesting counterpart of the '$\rho\pi$ puzzle' (see previous footnote) occurs also in the $\eta_c - \eta_c'$ system.

## Acknowledgements

We are greatly indebted to our technical collaborators, to the members of the CERN-SL Division for the excellent performance of the LEP collider, and to the funding agencies for their support in building and operating the DELPHI detector.
We acknowledge in particular the support of
Austrian Federal Ministry of Science and Traffics, GZ 616.364/2-III/2a/98,
FNRS–FWO, Belgium,
FINEP, CNPq, CAPES, FUJB and FAPERJ, Brazil,
Czech Ministry of Industry and Trade, GA CR 202/96/0450 and GA AVCR A1010521,
Danish Natural Research Council,
Commission of the European Communities (DG XII),
Direction des Sciences de la Matière, CEA, France,
Bundesministerium für Bildung, Wissenschaft, Forschung und Technologie, Germany,
General Secretariat for Research and Technology, Greece,
National Science Foundation (NWO) and Foundation for Research on Matter (FOM), The Netherlands,
Norwegian Research Council,
State Committee for Scientific Research, Poland, 2P03B06015, 2P03B03311 and SPUB/P03/178/98,
JNICT–Junta Nacional de Investigação Científica e Tecnológica, Portugal,
Vedecka grantova agentura MS SR, Slovakia, Nr. 95/5195/134,
Ministry of Science and Technology of the Republic of Slovenia,
CICYT, Spain, AEN96–1661 and AEN96-1681,
The Swedish Natural Science Research Council,
Particle Physics and Astronomy Research Council, UK,
Department of Energy, USA, DE–FG02–94ER40817, DE–FG02–87ER40344,
Alexander von Humboldt Stiftung, Bonn, Germany.

# References

[1] W. Kwong, J.L. Rosner and C. Quigg, Ann. Rev. Nucl. Sci. **37** (1987) 325–382;
F.J. Gilman, *"An Introduction to Charm and Heavy Quark Physics"*, Proceedings of the CCAST (World Laboratory) Symposium/Workshop – Charm Physics. Beijing, 1987. Ed. Y.Y. Tao Huang, p. 1–88.

[2] K.K. Seth, *"High Resolution Charmonium Spectroscopy By Antiproton Annihilation – Fermilab E760 and Beyond"*, Proceedings of AIP Conference – Few-Body Problems in Physics. Williamsburg, Va, May 1994. Ed. F. Gross, p. 248–274.

[3] C. Edwards et al. (Crystal Ball Coll.), Phys. Rev. Lett. **48** (1982) 70.

[4] Particle Data Group, Review of Particle Properties, Phys. Rev. **D54** (1996) 1.

[5] S. Kawabata, *"Physics in Two-Photon interactions"*, Proceedings of Joint International Lepton-Photon Symposium and Europhysics Conference on High Energy Physics. Geneva (1991). Eds. S. Hegarty et al., WSPC, Singapore, p.53–71.

[6] E.J. Eichten and C. Quigg, Phys. Rev. **D49** (1994) 5845 and private communications.

[7] J. Shigemitsu, Nucl. Phys. **B** (Proc. Suppl.) 53 (1997) 16;
P.B. Mackenzie, private communications.

[8] E.J. Eichten and C. Quigg, Phys. Rev. **D52** (1994) 1726.

[9] T. Barnes, T.E. Browder and S.F. Tuan, Phys. Lett. **B385** (1996) 391.

[10] T.A. Armstrong et al. (E760 Coll.), Phys. Rev. **D52** (1995) 4839.

[11] G. Zioulas, *"First Results from Fermilab E835"*, to be published in the Proceedings of International Symposium *HADRON 97*, BNL, August 1997;
K.K. Seth, *"Open Problems in Physics of Charmonium and QCD Exotics"*, to be published in the Proceedings of XXVII International Symposium on Multiparticle Dynamics, Frascati, September 1997.

[12] H. Albrecht et al. (ARGUS Coll.), Phys. Lett. **B338** (1994) 390;
O. Adriani et al. (L3 Coll.), Phys. Lett. **B318** (1993) 575;
D. Bisello et al. (DM2 Coll.), Nucl. Phys. **B350** (1991) 1;
W.-Y. Chen et al. (CLEO Coll.), Phys. Lett. **B243** (1990) 169;
V. Savinov and R. Fulton (for CLEO Coll.), contribution to the PHOTON '95 conference, Sheffield (1995). hep-ex/9507006.

[13] P. Aarnio et al. (DELPHI Coll.), Nucl. Instr. & Meth. **A303** (1991) 233.

[14] E.G. Anassontzis et al., Nucl. Instr. & Meth. **A323** (1992) 351.

[15] P. Abreu et al. (DELPHI Coll.), Nucl. Instr. & Meth. **A378** (1996) 57.

[16] P. Abreu et al. (DELPHI Coll.), Phys. Lett. **B401** (1997) 181.

[17] D. Bisello et al. (DM2 Coll.), Phys. Lett. **B200** (1988) 215.

[18] K.T. Chao, Y.F. Gu and S.F. Tuan, *"On Trigluonia in Charmonium Physics"*, BIHEP-TH-93-45, PUTP-93-24 and UH-511-790-94, December 1993;
S.F. Tuan, *"Hadronic Decay Puzzle in Charmonium Physics"*, UH-511-812-94, November 1994, to be published in the Proceedings of the 6th Annual Hadron Spectroscopy and Structure Colloquium (HSSC94), College Park, MD, August 1994.

[19] Y-Q. Chen and E. Braaten, Phys. Rev. Lett. **80** (1998) 5060.

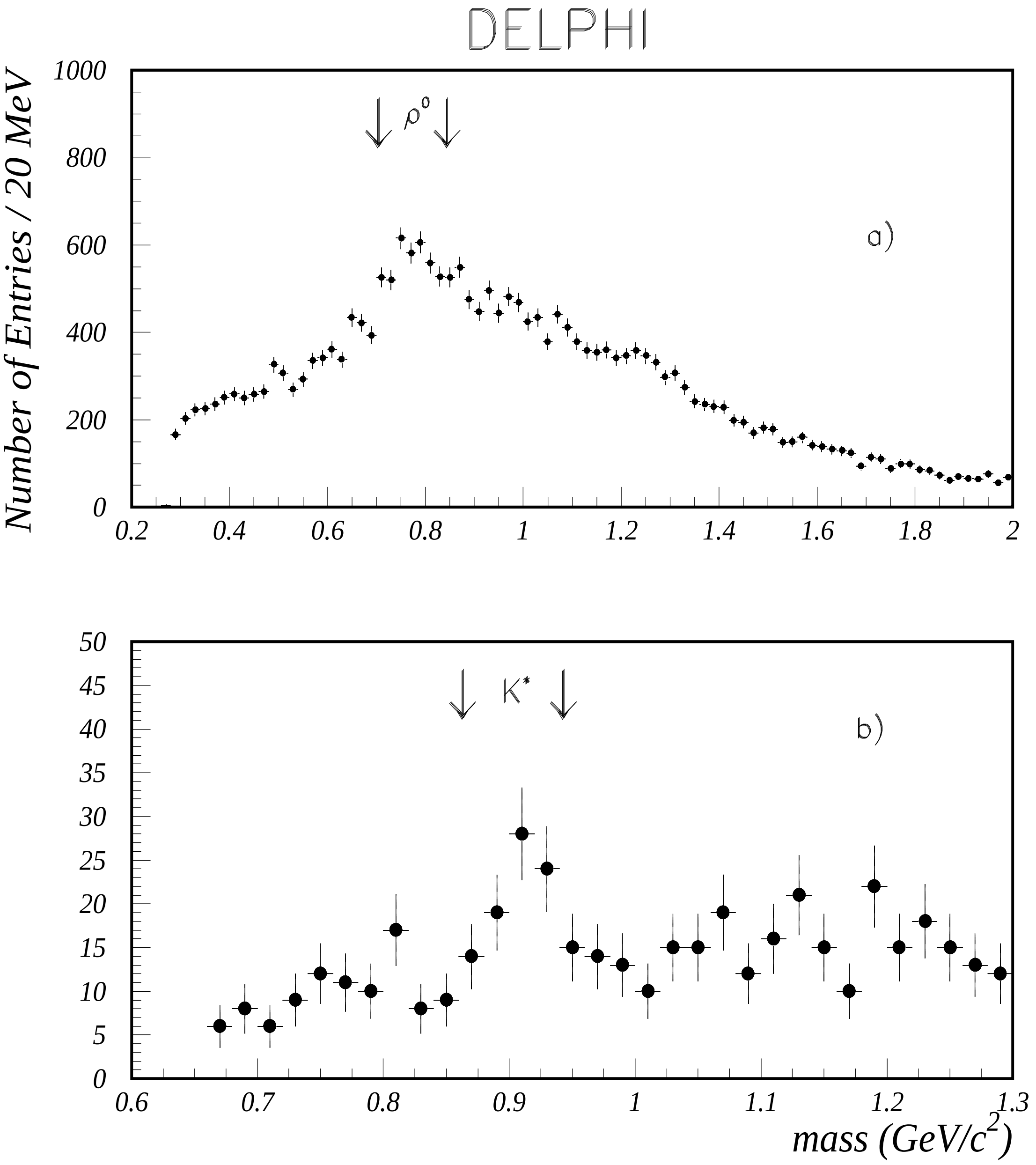

Figure 1: Invariant mass distributions (**a**) for $\pi^+\pi^-$ pairs in $\gamma\gamma \to \pi^+\pi^-\pi^+\pi^-$ events, (**b**) for $K^+\pi^-$ and $K^-\pi^+$ pairs in $\gamma\gamma \to K^+ K^-\pi^+\pi^-$ events.

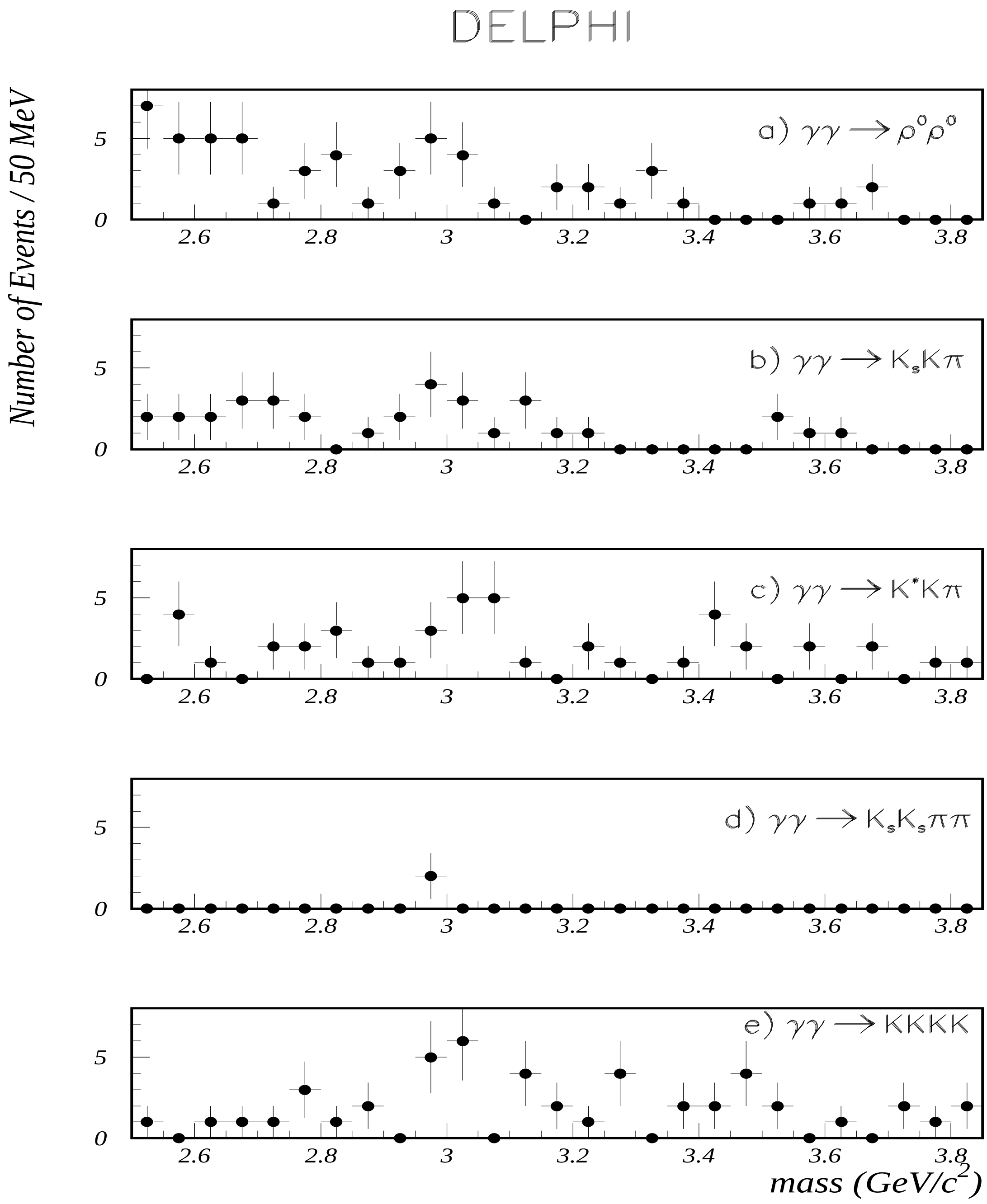

Figure 2: Invariant mass distributions of final state particles for the channels (**a**) $\gamma\gamma \to \rho^0\rho^0$, (**b**) $\gamma\gamma \to K_s^0 K^\pm \pi^\mp$, (**c**) $\gamma\gamma \to K^{*0} K^- \pi^+$ and $\gamma\gamma \to \bar{K}^{*0} K^+ \pi^-$, (**d**) $\gamma\gamma \to K_s^0 K_s^0 \pi^+ \pi^-$, (**e**) $\gamma\gamma \to K^+ K^- K^+ K^-$.

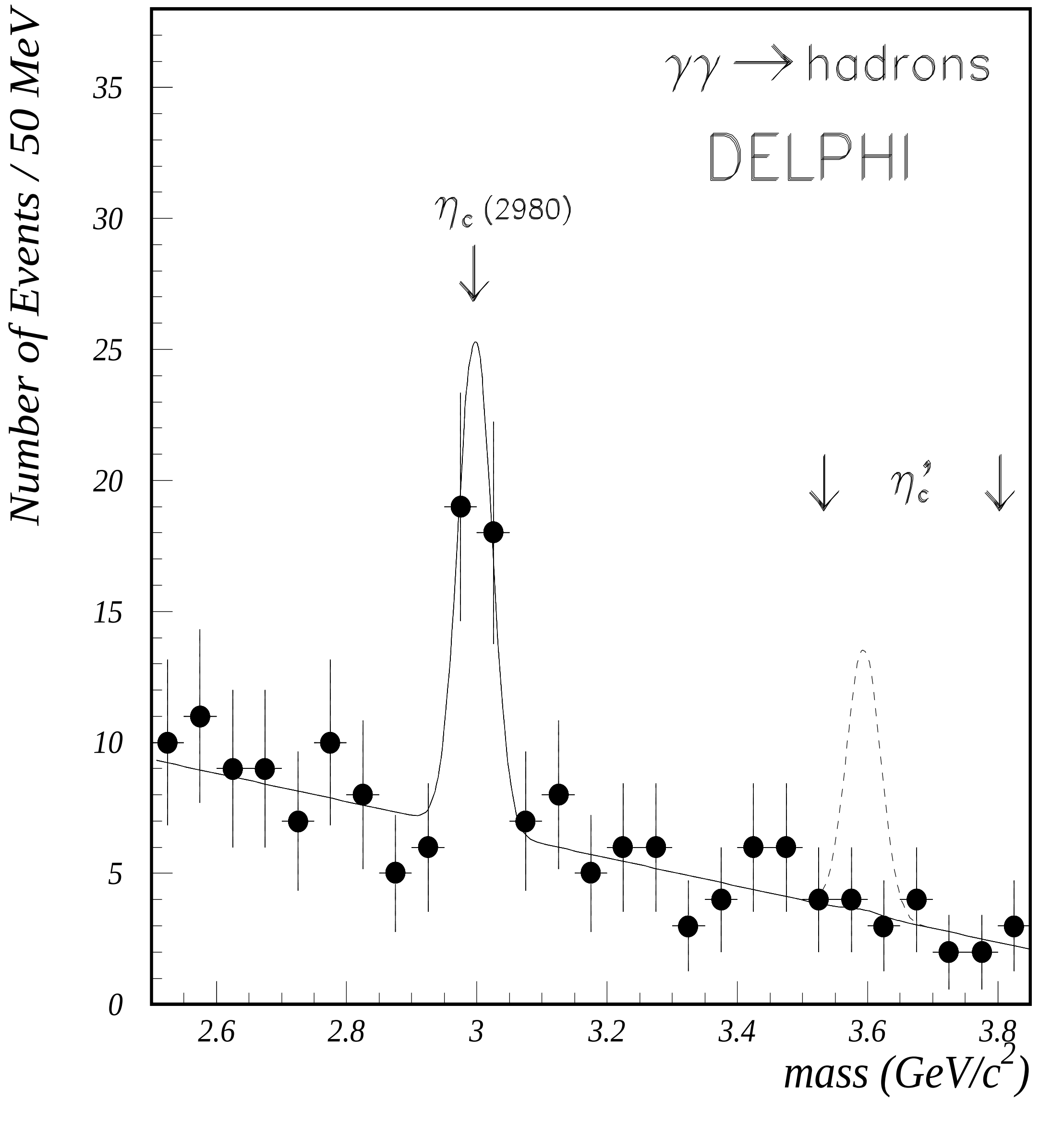

Figure 3: The sum of the invariant mass distributions of final state particles of the five $\gamma\gamma \to$ hadrons channels shown in Fig. 2. The full curve shows the result of the fit with the sum of two Gaussian functions for the $\eta_c$ and $\eta_c'$ signals and a linear term for the background contribution (see text). The dashed curve shows the expected signal of $\eta_c'$ production for $\Gamma_{\gamma\gamma}(\eta_c')/\Gamma_{\gamma\gamma}(\eta_c)$=0.75.